\documentclass[11pt,twoside]{article}

\usepackage{asp2014}

\newcommand{\hii}{H\textsc{ii}}

\aspSuppressVolSlug
\resetcounters

\begin{document}

\title{Detection and classification of radio sources with deep learning}

\author{Simone~Riggi,$^1$ Thomas~Cecconello,$^{1,2}$ Ugo Becciani,$^1$ and F. Vitello$^3$}
\affil{$^1$INAF - Osservatorio Astrofisico di Catania, Via Santa Sofia 78, 95123 Catania, Italy; \email{simone.riggi@inaf.it}}
\affil{$^2$DIEII, University of Catania, Viale Andrea Doria 6, 95125 Catania, Italy}
\affil{$^3$INAF - Istituto. di Radioastronomia, Via Gobetti 101, 40127 Bologna, Italy}

\paperauthor{Simone~Riggi}{simone.riggi@inaf.it}{0000-0001-6368-8330}{INAF}{Osservatorio Astrofisico di Catania}{Catania}{Catania}{95123}{Italy}
\paperauthor{Thomas~Cecconello}{thomas.cecconello@inaf.it}{0000-0001-6519-5011}{University of Catania}{Department of Electrical, Electronic and Computer Engineering}{Catania}{Catania}{95125}{Italy}
\paperauthor{Ugo~Becciani}{ugo.becciani@inaf.it}{0000-0002-4389-8688}{INAF}{Osservatorio Astrofisico di Catania}{Catania}{Catania}{95123}{Italy}
\paperauthor{Fabio~Vitello}{fabio.vitello@inaf.it}{0000-0003-2203-3797}{INAF}{Istituto. di Radioastronomia}{Catania}{Bologna}{40127}{Italy}

%\aindex{Riggi,~S.}
%\aindex{Cecconello,~T.}
%\aindex{Becciani,~U.}
%\aindex{Vitello,~F.}

\begin{abstract}
In this paper we present three different applications, based on deep learning methodologies, that we are developing to support the scientific analysis conducted within the ASKAP-EMU and MeerKAT radio surveys. One employs instance segmentation frameworks to detect compact and extended radio sources and imaging artefacts from radio continuum images. Another application uses gradient boosting decision trees and convolutional neural networks to classify compact sources into different astronomical classes using combined radio and infrared multi-band images. Finally, we discuss how self-supervised learning can be used to obtain valuable radio data representations for source detection, and classification studies. 
\end{abstract}

%\ssindex{astronomy!radio}
%\ssindex{observatories!ground-based!SKA}
%\ssindex{observatories!ground-based!ASKAP}
%\ssindex{observatories!ground-based!MeerKAT}
%\ssindex{astronomy!Galactic}
%\ssindex{algorithm!source finding}
%\ssindex{algorithm!machine learning!deep learning}
%\ssindex{software!image processing}
%\ssindex{techniques!convolutional neural networks (CNN)}
%\ssindex{classification}

\section{Introduction}
A new era in radio astronomy has begun as the SKA precursor telescopes started their planned survey programs. Among them, the Evolutionary Map of the Universe (EMU) \citep{Norris2011} of the Australian SKA Pathfinder (ASKAP, \citealt{Hotan2021}) started in 2022 to survey $\sim$75\% of the sky at 940 MHz with a target noise rms of $\sim$15 $\mu$Jy/beam and an angular resolution of $\sim$10". The EMU source cataloguing process will require an unprecedented degree of automation and knowledge extraction, as the expected number of detectable sources is $\sim$50 millions. So will be for other precursors and future SKA observations. In this context, machine learning (ML) can be a powerful resource for several data post-processing tasks, such as source detection, classification and anomaly discovery, shortening the time to deliver scientific results, and potentially enabling groundbreaking discoveries.\\In this paper we present three different application of deep learning methodologies on radio continuum data produced by SKA precursors: radio source finding with object detection frameworks (Section~\ref{sec:mrcnn}), compact  source type classification with multi-wavelength data (Section~\ref{sec:compact_source_classification}), and self-supervised learning of radio data for source detection and classification (Section~\ref{sec:ssl}).

\section{Source finding with deep neural networks}
\label{sec:mrcnn}
Algorithms used in traditional radio source finders are currently not suited to detect spurious or poor-imaged sources or extended radio sources that are composed by multiple disjoint regions. We are therefore evaluating object detection frameworks based on deep neural networks to tackle this problem.\\In \citep{RiggiMaskRCNN} we described 
a new source finder (\emph{caesar-mrcnn}) that uses the Mask R-CNN instance segmentation framework \citep{MaskRCNN} to detect objects of different classes in radio maps: \texttt{SPURIOUS}, 
\texttt{COMPACT}, \texttt{EXTENDED}, \texttt{EXTENDED-MULTISLAND}, \texttt{FLAGGED}. 
This tool was trained on a sample of $\sim$12.000 annotated images taken from the ASKAP EMU, VLA FIRST and ATCA Scorpio surveys, including $\sim$38.000 labelled objects (source class id and ground truth segmentation mask).\\
We have recently upgraded \emph{caesar-mrcnn} to TensorFlow 2.x, producing a new refactored version with an improved data pre-processing pipeline and with support for additional backbone models. Several training runs were then carried out with the same dataset to evaluate the performances against the original implementation. In Table~\ref{table_mrcnn} we report the source detection performances obtained over the test set with three different models. We found that the new implementation achieves slightly better source detection performances compared to the original version. The suboptimal performances on spurious and flagged sources are however retrieved also in the new version.\\
To improve the performances we need to either improve the dataset, the Mask R-CNN model, or eventually consider alternative frameworks. As the manual annotation effort required to increase the dataset size and improve its class unbalance has become unsustainable, we are considering latent diffusion models to conditionally generate synthetic radio images for desired classes and, at the same time, also their segmentation mask \citep{SortinoRADIFF}. 
To improve the Mask R-CNN model, we are exploring alternative data pre-processing strategies and self-supervised methods (see Section~\ref{sec:ssl}) for pre-training the model backbone. In \citep{SortinoSurvey} we carried out a survey of recent state-of-the-art frameworks for instance segmentation. YOLO v7 was identified as the best candidate to replace Mask R-CNN given the achieved accuracies and training speeds.

%\begin{itemize}
%\item \emph{Improving the dataset}: As the manual annotation effort required to increase the dataset size and improve its class unbalance has become unsustainable, we are considering latent diffusion models to conditionally generate synthetic radio images for desired classes and, at the same time, also their segmentation mask \citep{SortinoRADIFF};
%\item \emph{Improving Mask R-CNN model}: To improve the Mask R-CNN model, we are exploring alternative data pre-processing strategies and self-supervised methods (see Section~\ref{sec:ssl}) for pre-training the model backbone;
%\item \emph{Exploring alternative frameworks}: In \citep{SortinoSurvey} we carried out a survey of recent state-of-the-art frameworks for instance segmentation. YOLO v7 was identified as the best candidate to replace Mask R-CNN given the achieved accuracies and training speeds.
%\end{itemize}

\begin{table}[!ht]
\caption{Object detection metric $F=2\times\mathcal{C}\times\mathcal{R}/(\mathcal{C}+\mathcal{R})$ ($\mathcal{C}\equiv$ completeness, $\mathcal{R}\equiv$reliability), obtained with \emph{caesar-mrcnn} over the test sample with three different models: \emph{tf1-resnet101} (original TensorFlow 1.x implementation, \emph{ResNet101} backbone), \emph{tf2-resnet101} and \emph{tf2-resnet18} (new TensorFlow 2.x implementation, \emph{ResNet101} and \emph{ResNet18} backbones).}
\smallskip
\begin{center}
\small%
\begin{tabular}{lccccc}  % l = left, c = centered
\tableline
\noalign{\smallskip}
Model & \texttt{COMPACT} & \texttt{EXTENDED} & \texttt{EXTENDED-MULTI} & \texttt{SPURIOUS} & \texttt{FLAGGED} \\
\noalign{\smallskip}
\tableline
\noalign{\smallskip}
tf1-resnet101 & 70.6 & 83.7 & 75.6 & 44.0 & 81.9\\
tf2-resnet101 & 73.2 & 87.7 & 81.4 & 36.5 & 78.1\\
%tf2-resnet18 & 72.1 & 86.9 & 91.8 & 38.6 & 54.9\\
tf2-resnet18 & 75.4 & 79.8 & 79.8 & 36.9 & 54.9\\
\noalign{\smallskip}
\tableline%
\end{tabular}
\end{center}
\label{table_mrcnn}
\end{table}

\section{Compact radio source type classification}
\label{sec:compact_source_classification}
A large fraction of the sources extracted and catalogued in Galactic plane radio surveys are currently unclassified. Of these, more than 90\% are typically point-like or single-component resolved sources. In this scenario, new classification tools based on ML techniques and targeted for compact sources could therefore contribute to improve the census of Galactic sources.\\
To this aim we created a dataset of $\sim$20.000 previously classified compact sources, observed in different radio surveys: ASKAP EMU, VLA FIRST, GLOSTAR, THOR, MAGPIS, CORNISH. 
Seven source classes were considered: \texttt{RADIO-GALAXY}, \texttt{QSO}, \texttt{PN}, \texttt{\hii{}}, \texttt{PULSAR}, \texttt{YSO}, \texttt{STAR}. Radio source images were complemented with infrared data from the AllWISE survey (3.4, 4.6, 12, 22~$\mu$m), the GLIMPSE survey (8~$\mu$m), and the Hi-GAL survey (70~$\mu$m). The entire dataset was split in two major subsets on the basis of the available frequency bands for each source.
The 5-band dataset includes sources with radio and these IR frequencies:
3.4~$\mu$m, 4.6~$\mu$m, 12~$\mu$m, and 22~$\mu$m. The 7-bands dataset includes sources with radio and these IR frequencies: 3.4~$\mu$m, 4.6~$\mu$m, 8~$\mu$m, 12~$\mu$m, and 22~$\mu$m, 70~$\mu$m.\\
For sources in the 5-band (7-band) subset we computed 12 (18) radio-infrared feature parameters to be used in the classification analysis. The most sensitive parameters found were the radio-infrared colour indices $c_{i,j}$=$\log_{10}(F_j/F_i)$, where $F_{i}$ and $F_{j}$ are the source flux densities measured in band $i$ and $j$ with $\lambda_j$>$\lambda_i$. 
For both the 5-band (7-band) datasets we also measured the in-band radio spectral index for about $\sim$20\% ($\sim$40\%) of the sources. Finally, we have split both datasets in a train and test set, the latter including only sources observed with the ASKAP-EMU survey.\\A LightGBM classifier was trained on the extracted feature data and its classification metrics estimated on the test set for all the classes. We found that Galactic objects can be discriminated with high accuracy (above 90\%) against sources belonging to the extragalactic group. Classification performances largely vary among individual source classes. Extragalactic objects (radio galaxies, QSO) are best classified, with F1-scores exceeding 85\%. PNe, \hii{} regions, and pulsars are the second group of best classified objects, with F1-scores ranging from 60\% to 75\%. Poor performances are obtained on radio star group and YSOs, due to the limited sample size, object spectral type heterogeneity, and unreliable classification information reported in the reference catalogues.\\We also found a significant boost in performance ($\sim$10\%) for PNe, \hii{} regions, and pulsars when including additional infrared band data (8~$\mu$m, 70~$\mu$m) and the radio spectral index information in the analysis. On the 5-band dataset, the LightGBM classifier was also found to slightly outperform a convolutional neural network classifier directly trained on multichannel images.

\section{Self-supervised learning of radio data}
\label{sec:ssl}
Annotated datasets available for radio source analysis are typically very small and class-unbalanced, hampering the effectiveness of supervised approaches, particularly when training large models. Self-supervised learning \citep{Liu2023} has recently emerged as a powerful methodology to deal with some of the aforementioned problems. A self-supervised framework can be in fact trained on large samples of unlabelled data and the resulting model backbone can be used to obtain representation features of new input data for data inspection and unsupervised analysis. Moreover, this model can be also fine-tuned over labelled data to carry out various down-stream tasks, e.g. source detection and classification.\\
We have trained SimCLR \citep{SimCLR}, a popular self-supervised contrastive learning framework, over each one of these unlabelled ASKAP-EMU survey data: 1) $\sim$170.000 radio images of pixel size 256$\times$256, randomly extracted from the survey; 2) $\sim$17.000 radio images centred on resolved sources with different morphologies, including diffuse sources. A third model was also trained over both datasets in a two-step sequence. For all models we used a \emph{ResNet18} encoder and a 2-layer projector, and applied these augmentation transforms: random cropping, colour jitter, rotation, and flipping. Trained models were used to perform various down-stream task studies.\\We studied how the learned representation and the self-supervised model perform on a source classification problem using the \emph{Radio Galaxy Zoo} (RGZ) dataset \citep{RGZ}. 
We extracted $\sim$80.000 images from the VLA FIRST survey around RGZ sources that have been labelled into 6 different morphological classes on the basis of the observed number of components (C) and peaks (P):
\texttt{1C-1P}, \texttt{1C-2P}, \texttt{1C-3P}, \texttt{2C-2P}, \texttt{2C-3P}, \texttt{3C-3P}. A source classifier with standard architecture (a \emph{ResNet18} backbone as in SimCLR model, and a single-layer classification head) was trained on a class-balanced dataset including 1000 (600) images per class in the train (test) set.
We first froze the model backbone, fixing its weights to those obtained in the trained SimCLR models, and trained only the classification head. The classification F1-scores obtained by different self-supervised pre-trained models on the test set were compared against a baseline model pre-trained on the \emph{ImageNet} dataset \citep{ImageNet}. All self-supervised models were found to outperform the baseline, particularly the third model that achieved $\sim$10\% higher scores for all classes.\\
We also fine-tuned the source classifier and studied the accuracies reached as a function of the number of labelled data available in the training set, to investigate how the model performs in the low-label regime. We compared two models: one trained from scratch, and the other with backbone weights initialized to the best SimCLR model backbone weights. We found that the fully supervised model becomes almost untrainable (scores<0.45) below a fraction of $\sim$10\% of the original train dataset. On the other hand, self-supervised pre-trained model reaches scores $\sim$0.65 with only 2\% of labels using a considerably smaller number of train epochs. Above the 10\% label fraction threshold, the fully supervised model outperforms by a few percent the other model, but with a significantly larger cost in training times. 

\acknowledgements This work received funding from the INAF CIRASA and SCIARADA projects, and from the Italian MUR project CN\_00000013 "Italian Research Center on High-Performance Computing, Big Data and Quantum Computing", Spoke 3 Astrophysics and Cosmos Observations, National Recovery and Resilience Plan.

\bibliography{C402}%

\end{document}